\documentclass[journal=jacsat,manuscript=article,layout=twocolumn]{achemso}
\usepackage[T1]{fontenc} 
\usepackage{float}
\usepackage{mciteplus}
\usepackage{subcaption}
\usepackage{chemformula} 
\author{Chen-Yen Lai}
\affiliation{Theoretical Division, Los Alamos National Laboratory, Los Alamos, New Mexico 87545, USA}
\alsoaffiliation{Center for Integrated Nanotechnologies, Los Alamos National Laboratory, Los Alamos, New Mexico 87545, USA}
\email{chengyanlai@gmail.com}
\author{D. A. Yarotski}
\affiliation{Center for Integrated Nanotechnologies, Los Alamos National Laboratory, Los Alamos, New Mexico 87545, USA}
\author{Jian-Xin Zhu}
\affiliation{Theoretical Division, Los Alamos National Laboratory, Los Alamos, New Mexico 87545, USA}
\alsoaffiliation{Center for Integrated Nanotechnologies, Los Alamos National Laboratory, Los Alamos, New Mexico 87545, USA}

\title[An \textsf{achemso} demo]
  {Time-resolved Photoluminescence in Terahertz-driven Hybrid Systems of Plasmons and Excitons}
\abbreviations{}
\begin{document}



\begin{abstract}
  Ultrafast pump-probe technique is a powerful tool to understand and manipulate properties of materials for designing novel quantum devices.
  An intense, single cycle terahertz pulse can change the intrinsic properties of semiconductor quantum dots to have different luminescence.
  In a hybrid system of plasmon and exciton, the coherence and coupling between these two degrees of freedom play an important role on their optical properties.
  Therefore, we consider a terahertz pump optical probe experiment in the hybrid systems where the terahertz pump pulse couples to the exciton degrees of freedom on the quantum dot.
  The time resolved photoluminescence of the hybrid system shows that the response of the characteristic frequency shifts according to the overlap between the pump and probe pulses.
  Furthermore, the resonance between the exciton and plasmons can be induced by the terahertz pump pulse in some parameter regimes.
  Our results show the terahertz driven hybrid system can be a versatile tool for manipulating the material properties and open a new route to design modern optical devices.
\end{abstract}

\maketitle

Recently, ultrafast techniques have drawn a lot of attention because of the capability to access novel quantum states of matter from various different pump-probe configurations.~\cite{zhang2014,basov2017}
For example, the time-resolved, angle-resolved photoemmision spectroscopy can provide information on single particle spectra~\cite{perfetti2007,tao2010,brown2019};
time-resolved Raman scattering for the particle-hole excitations~\cite{tao2012,wang2018a,shinjo2018}; time-resolved x-ray photoluminescence spectroscopy for the dynamics from core hole effect~\cite{lai2019a,tan2012,vanaken2002}; and time-resolved resonant inelastic x-ray scattering for the dynamics of transient excitations.~\cite{kim2004,ament2011a,ament2011,benjamin2014,nocera2018b,chen2019b,kang2019a,dean2016}
Beside probing the dynamics of excitations, state-of-the-art pump-probe techniques also provide new direction in manipulating properties of materials and open up a new class of photonic devices with controllable functionalities.~\cite{hayat2012,xu2008,bowlan2017}

An intense single cycle terahertz frequency pump pulse can provide multi-MV/cm field strengths to access nonlinear spectroscopy applications~\cite{liu2012} and control the structural dynamics.~\cite{bowlan2017}
Recent studies~\cite{hoffmann2010,pein2017} use the terahertz electric field enhancement to manipulate the properties of quantum dots (QD) by measuring the reflectivity and photoluminescence (PL) on gold microslits,
which demonstrate a path way to electro-optic modulation from the interaction between the applied electric field and the excitonic degrees of freedom of the QD.
On the other hand, the plasmon degrees of freedom on the nanoparticle (NP) can couple effectively to the exciton on the QD.~\cite{hollingsworth2015}
In hybrid systems, one can modify the spontaneous emision through assembly and synthesis~\cite{hollingsworth2015} to achive higher intensity and tunability near resonance.
The anisotropic nature of the hybrid systems also bring more interesting strucrture~\cite{lai2019} in the PL
which makes the hybrid system a potential candidate for designing photonic devices
such as nanoattena~\cite{savasta2010,marinica2013} and quantum information devices.~\cite{pelton2010}

In this article, a quantum mechanical model we proposed previuously~\cite{lai2019} is used to describe the exciton plasmon hybrid system consisting of a single quantum dot and nanoparticle.
We calculate the nonequilibrium PL of the hybrid system under the influence of external terahertz pump field.
The probe field with a time delay can overlap with the pump field which results in a time-resolved PL (tr-PL).
The nonequilibrium spectrum is calculated in time dependent perturbation theory where the probe field is treated as a perturbation in linear response and the formula shares similar structure with the time-resolved, angle-resolved photoemmision spectroscopy~\cite{freericks2009} and time resolved x-ray absorption spectroscopy.~\cite{wang2019}
The results show the frequency shift of PL peaks according to the overlap between the pump and probe fields which reveals the manipulation of the optical properties of the hybrid systems.
The frequency shift is proportional to the instantaneous intensity of the pump pulse also known as ac-Stark effect.
In addition, our model calculation also demonstrates that there is a terahertz driven resonance in the hybrid system in the proper parameter regimes.
Our studies of tr-PL show that the terahertz driven exciton-plasmon hybrid systems can extend the applicability and functionality as a potential photonic device.

We model the hybrid system of NP-QD with the Hamiltonian~\cite{lai2019}
\begin{eqnarray}
  H_{\text{system}} = E_dd^\dagger d + E_cc^\dagger c - \Delta_{dc}(d^\dagger + d)(c^\dagger + c)\;.
\end{eqnarray}
Here, $c^\dagger(c)$ is the creation (annihilation) operator for excitons on the QD and $d^\dagger(d)$ is the creation (annihilation) operator for plasmons on the NP.
The internal energy ($E_{c(d)}$) levels are related to the physical sizes and materials of the NP and QD.~\cite{karimi2018,elward2013}
On the QD, only single exciton excitation is considered and there can be many plasmons occupied on the NP.
The most important coupling between excitons and plasmons in the first order is determined by dipole-dipole interaction~\cite{nordlander1986,manjavacas2011} and can be anisotropic.~\cite{lai2019}
Here, we only consider a single QD and one NP with isotropic coupling which can be achieved in experiment by aligning the polarization of the probe field parallel to the symmetry axis.
Thus, the parameter $\Delta_\text{dc}$ depends on their intrinsic properties and the distance between the NP and the QD.

\begin{figure}[t]
  \begin{center}
    \includegraphics[width=\columnwidth]{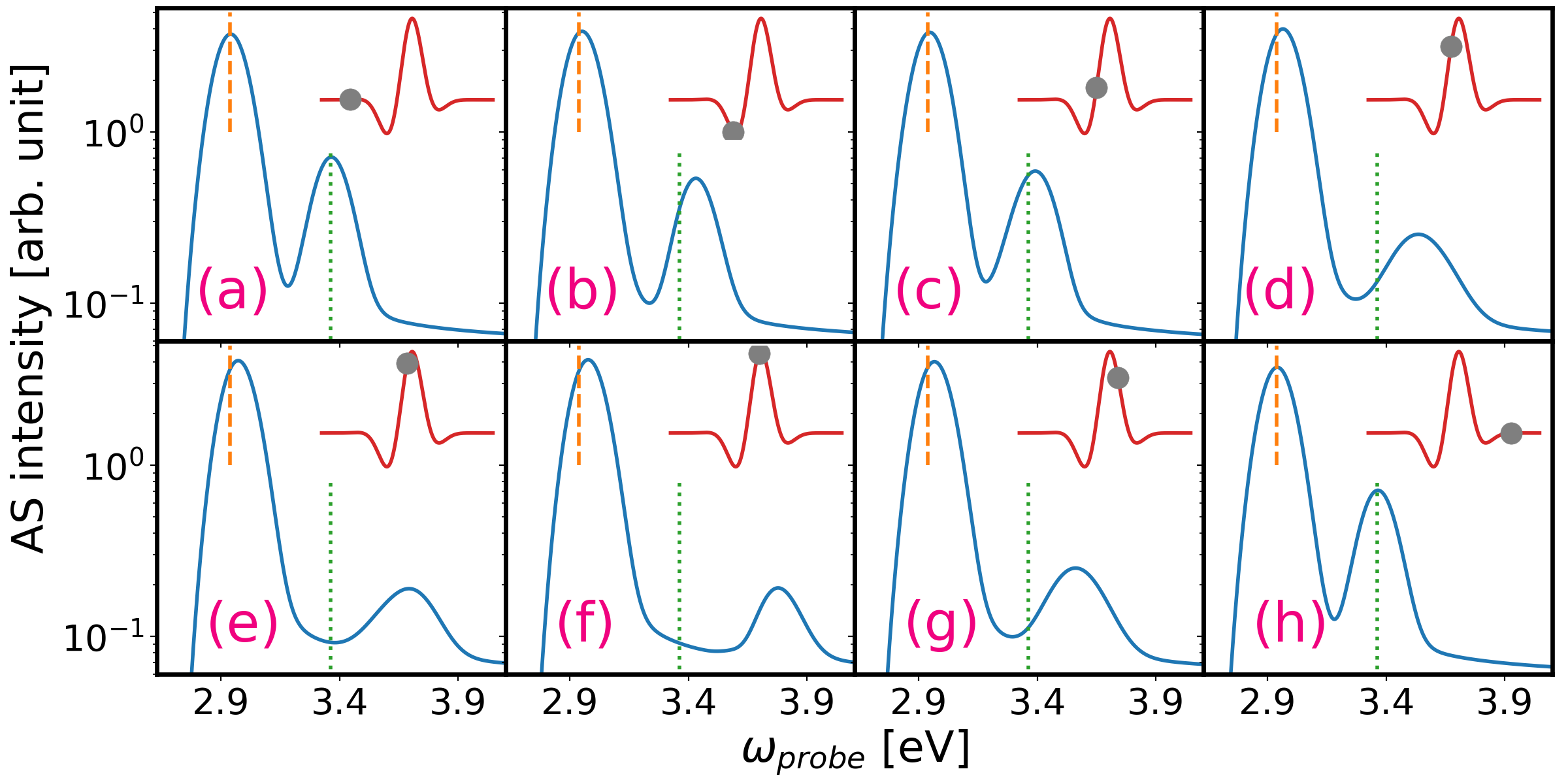}
    \caption{
      (a) The intensity of tr-PL of the hybrid system with different time delays (a) $-414$, (b) $-165$, (c) $-62$, (d) $-21$, (e) $0$, (f) $21$, (g) $93$, and (h) $414$ fs.
      The dashed (dotted) line in is frequency of the photoluminescence peak $\omega_d$ ($\omega_c$) at equilibrium from Eq.~\eqref{eq:eqm}.
      The insets show the temporal profile of the pump pulse and the circle marks the center of the probe pulse ($t_d$).
      The system has parameters of $E_d\!=\!3eV$, $E_c\!=\!3.3eV$, and $\Delta_\text{dc}\!=\!150meV$.
      The pump pulse has intensity $h_0\!=\!1$ and the probe pulse width is set to $\sigma_\text{probe}\!=\!41.4$ fs.}
    \label{fig:1ThzEc33}
  \end{center}
\end{figure}

In general, the pulse (both pump and probe) will couple to the NP and QD degrees of freedom through the dipole transition operator,
\begin{eqnarray}\label{eq:pulse}
  H_{k}(t) = h_0^{(k)}s_k(t,t_d)e^{i\omega_k t}(A^\dagger_k + A_k)\hat{e}_k\;,
\end{eqnarray}
where $k\!=\!$ pump or probe, $h_0$ is the intensity and $\hat{e}_k$ is the polarization.
Here, the $s_k(t,t_d)$ is the Gaussian envelope function of the pulse centered at $t_d$,
\begin{eqnarray}
s_k(t,t_d) = \frac{1}{\sqrt{2\pi}\sigma_{k}} \exp[-(t-t_d)^2/2\sigma^2_{k}]\;.
\end{eqnarray}
Usually, the polarizations of the pump and probe pulse are perpendicular to each other in experiments.
Here, the operator $A_k$ triggers the dipole transitions,
\begin{equation}
  A_k = M^{\bf e}_{d,k} d +  M^{\bf e}_{c,k} c
\end{equation}
where $M^{\bf e}\!\in\!C$ is the associate matrix element with photon polarization ${\bf e}$.
Generally, $M^{\bf e}$ can be time-dependent in non-equilibrium dynamics,
but it is treated as a time-independent parameter in this work for simplicity.
In order to mimic the experimental probe which detects the photon intensity with the incident frequency, the process is usually related to certain expectation value of the probe observable.
Here, we use time dependent perturbation theory.
Typically, the pump pulse $H_\text{pump}$ is stronger than the probe pulse $H_\text{probe}$, and one can treat the probe pulse as a perturbation.
Similar to the tr-ARPES and tr-XAS process~\cite{freericks2009,wang2019}, the tr-PL can be captured by
\begin{eqnarray}\label{eq:troas-final}
  I^\text{PL}(\omega_\text{probe},t_d)\approx\int\int dt_2dt_1e^{i\omega_\text{probe}(t_2-t_1)}\times&&\nonumber\\
  s_\text{probe}(t_1,t_d)s_\text{probe}(t_2,t_d)C(t_1,t_2)&&
\end{eqnarray}
where $C(t_1,t_2)\!=\!\langle (A^\dagger_\text{probe}(t_2) \!+\! A_\text{probe}(t_2)) (A_\text{probe}(t_1) \!+\! A^\dagger_\text{probe}(t_1))\rangle$ is a nonlocal time correlation function and $A_\text{probe}(t)\!=\!e^{i\mathcal{H}(t)}A_\text{probe}e^{-i\mathcal{H}(t)}$ with $\mathcal{H}(t)\!=\!H_\text{system}\!+\!H_\text{pump}(t)$.
The derivation of the above formula is given in the supplementary material.
The main task is to simulate the correlation function $C(t_1,t_2)$ in the real time evolution of the wave function.
Since the pump pulse couples different particle number sector, we include the Hilbert space up to ten total particles (one exciton plus nine plasmons or ten plasmons without exciton).
Under current consideration, the calculated spectrum is beyond rotating wave approximation and accounted for contribution from all included particle sectors.
In our simulation, usually only four total particles are needed since the terahertz pulse is off resonance with the energy scale of both QD and NP and the pulse intensity ($h_0$) is not set to be strong compared to the energy scales.
We set pump pulse with frequency of $\omega_\text{pump}\!=\!1$ terahertz, width of $\sigma_\text{pump}\!\approx\!250$ fs, and a phase shift $\phi\!=\!-0.2$, shown in the insets of Fig.~\ref{fig:1ThzEc33}, throughout entire article.

\begin{figure}[t]
  \begin{center}
    \includegraphics[width=0.9\columnwidth]{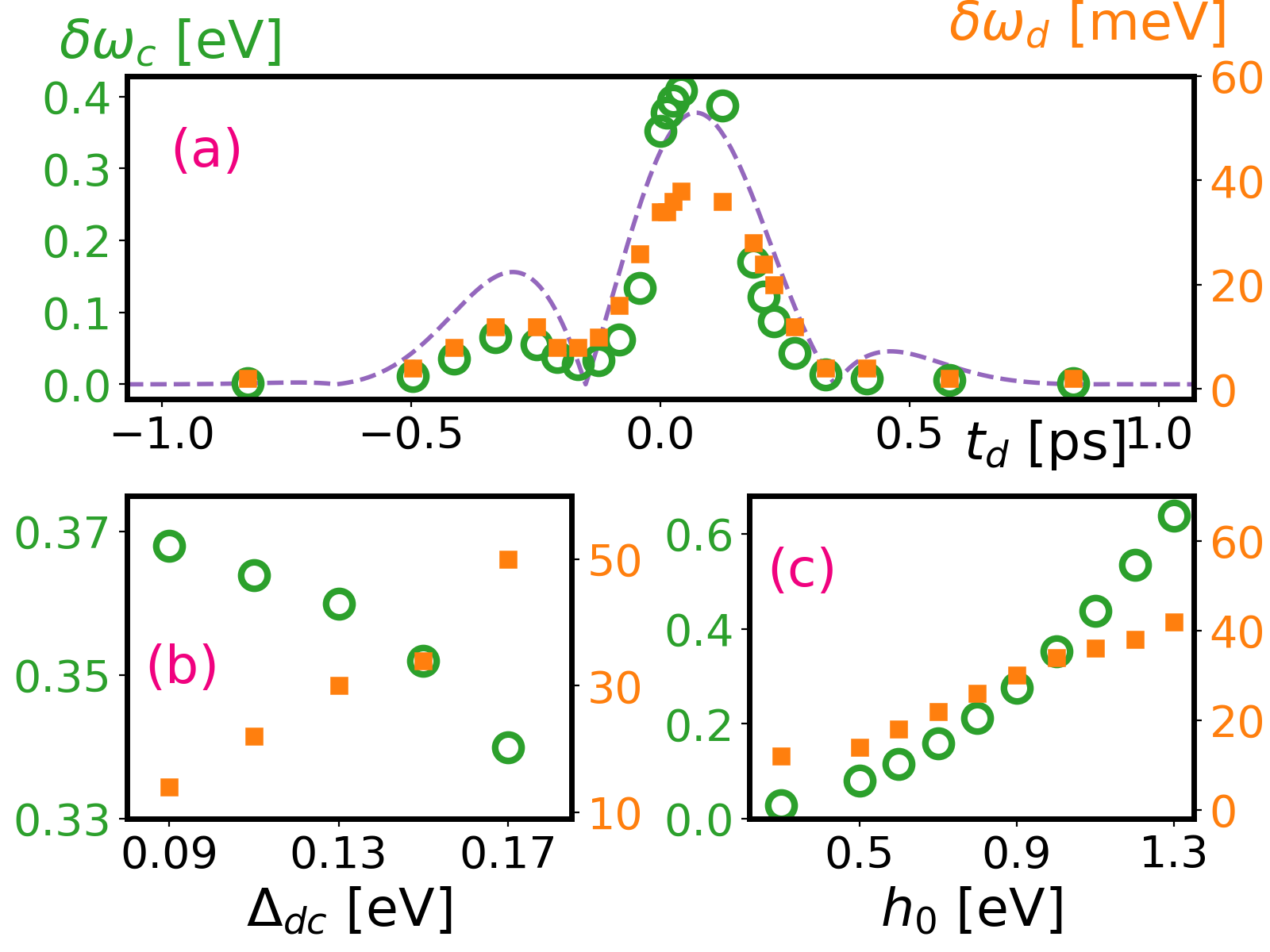}
    \caption{
      The energy shifts of both major peaks, $\delta\omega_c$ (open circle) and $\delta\omega_d$ (filled square), versus time delay (a), $\delta_\text{dc}$ (b) and $h_0$ (c).
      The dashed line in (a) shows the intensity of the terahertz pump pulse over time.
      The system has parameters of $E_d\!=\!3eV$, $E_c\!=\!3.3eV$, and $\Delta_\text{dc}\!=\!150meV$ in (a) and (c).
      The pump pulse has intensity $h_0\!=\!1$ in (a) and (b) and the probe pulse width is set to $\sigma_\text{probe}\!=\!66.2$ fs.
      }
    \label{fig:1ThzdOmega}
  \end{center}
\end{figure}

We focus on the regime where the terahertz pump pulse only couples to the QD as $M_{c,\text{pump}}\!\gg\!M_{d,\text{pump}}$ and the probe pulse excites the excitation on the NP as $M_{d,\text{probe}}\!\gg\!M_{c,\text{probe}}$.
Figure~\ref{fig:1ThzEc33} shows the tr-PL of the hybrid system under different time delays.
In equilibrium, the main feature of the spectrum has two photoluminescence peaks which locate near the energy scales~\cite{lai2019}
\begin{equation}\label{eq:eqm}
  \omega_{d,c}=\frac{1}{2}\left[E_d+E_c\pm\sqrt{(E_d-E_c)^2+4\Delta^2_\text{dc}}\right]\;.
\end{equation}
If the probe pulse is sent before the pump pulse starts to amplify, for example $t_d\!\approx\!-414$ fs, the spectrum is similar to the one in equilibrium as shown in Fig.~\ref{fig:1ThzEc33}(a) where the frequency of both photoluminescence peaks agrees with the equilibrium results which are marked as two vertical lines.
For large time delay where the pump pulse deminishes, both photoluminescence peaks recover to the similar energy scale as the equilibrium.
These results meet our expectation that the wavefucntion is likely to have high fidelity after the pump pulse deminishes since it is off resonance with the QD and its strength is also not strong.
Although the pump pulse is off resonance with the energy scale of the QD, the frequency of the photoluminescence peak shifts if the probe pulse overlaps with the pump pulse while the varying time delay $t_d$ of the probe pulse.
The photoluminescence peak $\omega_c$ has a major shift, $400$ meV if probed near the maximum intensity of the pump pulse, due to this overlap and the coupling to the pump pulse.
Furthermore, varying the time delays, the shifts of both peaks follow the instantaneous intensity of the pump pulse as shown in Fig.~\ref{fig:1ThzdOmega}(a).
The results show that the terahertz pump pulse affects the internal energy structure of the QD and the spectra depends on the instantaneous intensity of the pump pulse which indicates the ac-Stark effect caused by the terahertz driving fields.
This effect should be measurable by spatiotemporally overlapping the pump pulse and the probe pulse in the hybrid system.
Although the pump pulse does not directly couple to the NP, the energy of the photoluminescence peak $\omega_d$ also shifts accordingly which is affected by the coupling between NP and QD.
As the couplings change, the frequency shift of both major peak behaves differently as shown in Fig.~\ref{fig:1ThzdOmega}(b).
While the frequency shift of the $\omega_d$ increases as the couplings become stronger, the frequency shift of $\omega_c$ peak decreases.
However, the shifts of frequency do not strongly depend on the coupling as they only chenge around $40$ meV as the coupling nearly doubles.
As one chenges the intensity of the pump pulse, the scalings of the frequency shifts are different for both major peaks as shown in Fig.~\ref{fig:1ThzdOmega}(c).
While the frequency shift of the $\omega_c$ peak increases as $h_0^2$, the intensity of the peak deminishes (not shown).
This can be understood as the effective internal energy scale becomes large due to the pump pulse fluence and the larger difference between effective $E_d$ and $E_c$ results in smaller intensity in the $\omega_c$ peak which agrees with the conclusions from equilibrium.~\cite{lai2019}
For the frequency shift of the $\omega_d$ peak, it scales linearly between $h_0\!\sim\!0.5\!-\!1.0$ eV.
It is worth noting that the results presented here assume that the probe pulse has strong coupling to the plasmons on NP and not excitons on QD.
One can expect that the scaling and the relation can change if the probe pulse couples to the excitons more strongly than the plasmons.

\begin{figure}[t]
  \begin{center}
    \includegraphics[width=\columnwidth]{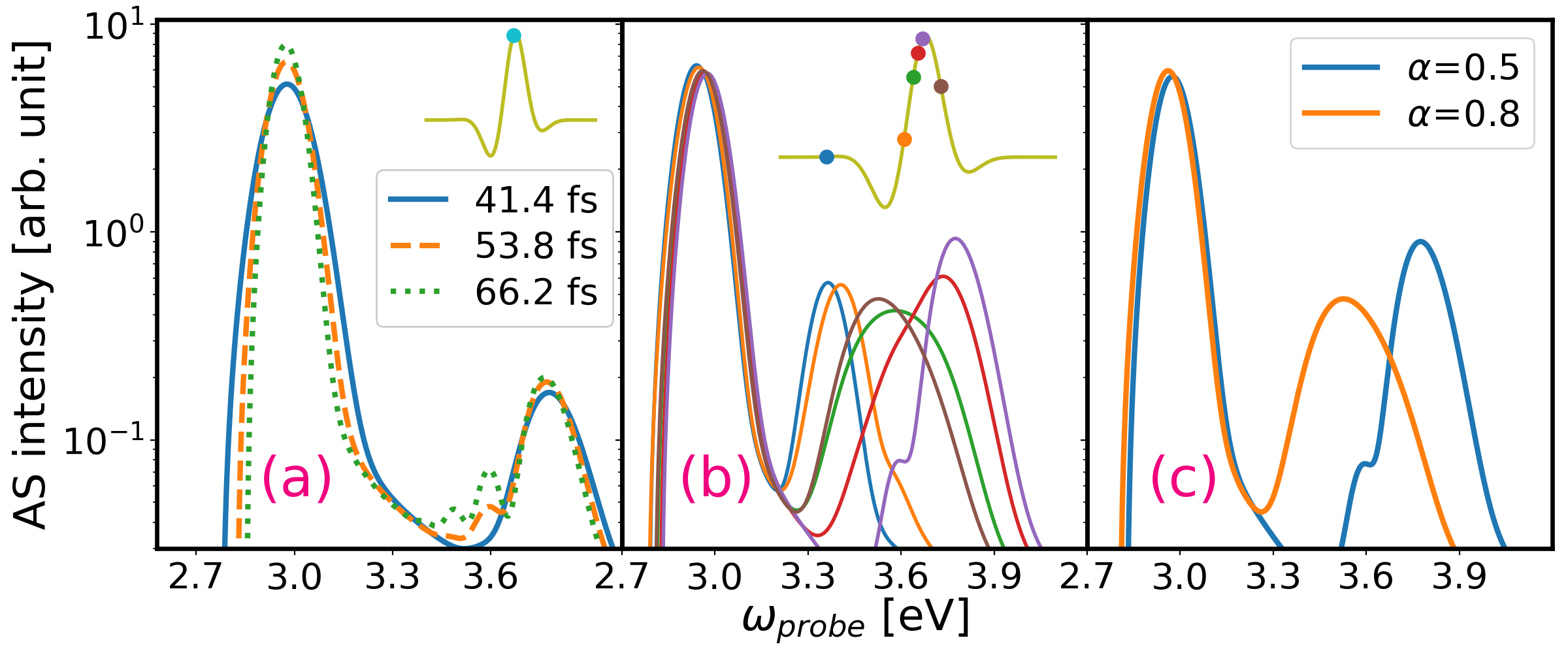}
    \caption{
      (a) The tr-PL spectrum under different probe pulse width $\sigma_\text{probe}$.
      The inset shows the profile of pump pulse and the time delay is marked.
      (b) The tr-PL spectrum with the probe pulse parameter $\alpha\!=\!0.5$ with different time delay which is marked in the inset with corresponding color.
      (c) With the same time delay, two different coupling ratio is compared.
      The system has parameters of $E_d\!=\!3eV$, $E_c\!=\!3.3eV$, and $\Delta_\text{dc}\!=\!150meV$.
      The pump pulse has intensity $h_0\!=\!1$ and the probe pulse width is set to $\sigma_\text{probe}\!=\!66.2$fs.
    }
    \label{fig:1THzSigmaRatio}
  \end{center}
\end{figure}

Beside the shifts of the photoluminescence peaks, there is another emergent peak if the width of probe pulse is wide enough and the time delay is around the maximum of the pump pulse.
Figure~\ref{fig:1THzSigmaRatio}(a) shows the comparison of the different width of the probe pulse.
The location of $\omega_c$ is shifted to $3.77$ eV and satellite peak(s) around $3.5-3.6$ eV.
This emergent peak is only visible when the width of the probe pulse is large.
For a narrow probe pulse, the spectrum is rather broaden and only two major dominant peaks are visible.
For a wilder probe pulse, the overlap between the pump and probe pulse extents and the spectrum reveals more energy scales that is driven by the pump pulse.
Considering different coupling ratio on the probe pulse ($\alpha\!=\!\vert M_{d,\text{probe}}/M_{c,\text{probe}}\vert$), the results give similar conclusions qualitatively as shown in Fig.~\ref{fig:1THzSigmaRatio}(b).
Here, the maximum energy shift still happens when the center of probe pulse is near the maximum of the pump pulse.
As the ratio decreases, which means the probe pulse couples to QD stronger, the energy shift of the $\omega_c$ peak becomes more pronounced as shown in Fig.~\ref{fig:1THzSigmaRatio}(c).

\begin{figure}[t]
  \begin{center}
    \includegraphics[width=\columnwidth]{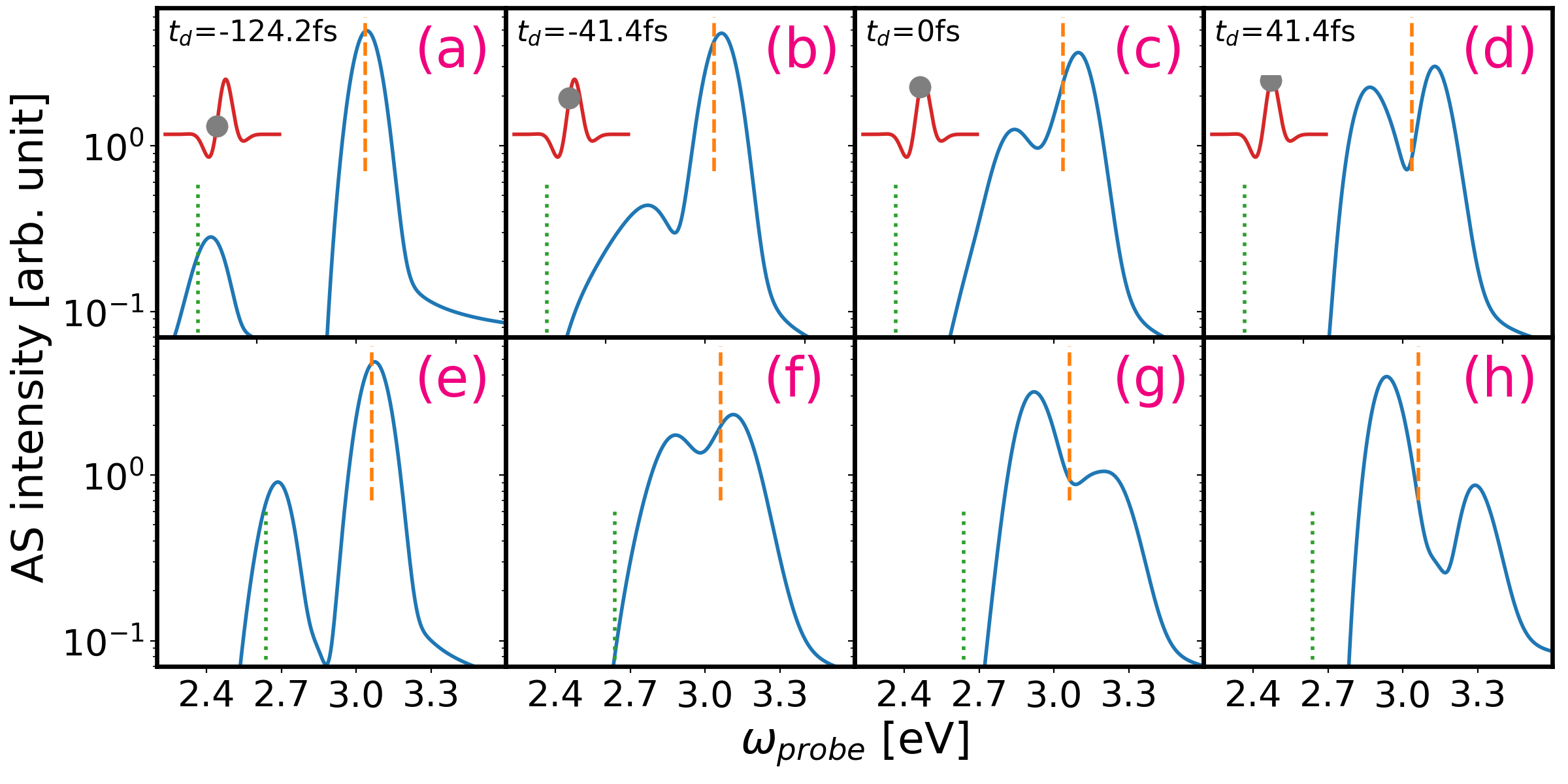}
    \caption{
      The spectrum with $E_c\!=\!2.4$ev (a)-(d) and $2.7$eV (e)-(h) is shown with different time delays shown in the insets.
      In equilibrium, the photoluminescence peak $\omega_d$ ($\omega_c$) is marked with vertical dashed (dotted) lines.
      The insets in (a)-(d) shows the pump pulse profile and the center of probe pulse.
      The system has parameters of $E_d\!=\!3eV$ and $\Delta_\text{dc}\!=\!150meV$.
      The pump pulse has intensity $h_0\!=\!1$ and the probe pulse width is set to $\sigma_\text{probe}\!=\!66.2$fs.
    }
    \label{fig:1THzEc24Ec27}
  \end{center}
\end{figure}

In the previous section, the tr-PL shows overall blue shift of both photoluminescence peaks compared to the equilibrium spectrum and the overall shift of the $\omega_c$ peak is larger than the other.
In equilibrium, the spectrum shows the resonance between plasmons and excitons when their intrinsic energy is very closed to one another~\cite{lai2019} where the intensity of both photoluminescence peaks is equal despite the different line shape.
Here, we consider the case where the energy of the QD is smaller than the one of NP $E_c\!<\!E_d$ which the $\omega_c$ peak will have smaller energy than the $\omega_d$ one.
In experiment, by using different materials, one can engineer the QD and NP which has different internal energy level and achieve the condition.~\cite{hollingsworth2015}
Figure~\ref{fig:1THzEc24Ec27} show the spectrum with two different $E_c\!=\!2.4$ and $2.7$eV while the energy of the NP remains the same at $3$eV.
In the early probe in Figs.~\ref{fig:1THzEc24Ec27}(a) and~\ref{fig:1THzEc24Ec27}(e), the spectrum shows similar results where both peak has blue shift and the $\omega_c$ peak has larger energy shift than the $\omega_d$ peak.
As the center of probe pulse moves toward the maximum of the pump pulse, the energy of $\omega_c$ peak move closer to the $\omega_d$ peak and its intensity also becomes stronger as shown in Figs.~\ref{fig:1THzEc24Ec27}(b) and~\ref{fig:1THzEc24Ec27}(c).
For $E_c\!=\!2.4$ev shown in Fig.~\ref{fig:1THzEc24Ec27}(d), the probe pulse center is near the maximum of pump pulse, both peak have nearly the same intensity but slightly different line shape.
This spectrum is very similar the resonance case in equilibrium when the internal energy scale of NP and QD are closed to each other.
The resonance predicted here, however, is driven by the terahertz pump pulse due to different response from exciton and plasmons.
Similarly, for $E_c\!=\!2.7$eV, the resonance happens before the center of the probe pulse is at the maximum of pump pulse as shown in Fig.~\ref{fig:1THzEc24Ec27}(f).
Interestingly, as the center keeps move toward the maximum of the pump pulse, the $\omega_c$ peak will appear in a higher energy than the $\omega_d$ peak.
Also, at the same time, the $\omega_d$ peak results in having a red shift instead of a blue shift.
Both effects observed here shows that the terahertz pump pulse indeed changes the internal energy scale of the QD which is the degree od freedom the pulse couples to.

We study a QD-NP hybrid system under the influence of a terahertz pump pulse by calculating its time resolved PL with various time delays.
The frequencies of the photoluminescence peaks are shifted according to the overlap between the pump pulse and probe pulse.
More importantly, the energy shifts are propotional to the instataneous intensity of the pump pulse which is a consequence of modifed properties of the QD as ac-Stark effect.
Also, the frequency shifts behaves differently as the couplings between the QD and NP change, and the relation also agrees with the conclusion from equilibrium by consideriong the effective eternal energy scale due to the the terahertz pump pulse.
Under the assumption of probe pulse couples strongly to the plasmon, the frequency shift scale quadraticly (linearly) as the intensity of the pump pulse for $\omega_{c(d)}$ peak.
Furthermore, if the QD has smaller internal energy than the NP, our results show that the terahertz pump pulse can drive the system into a resonance where both photoluminescence peak have nearly equal intensity and the frequency difference is closed to the coupling between plasmon and exciton.
This terahertz driven resonance is an important feature for designing novel photonic devices and various applications.
In experiment, the QD and NP particles have roughly internal energy as several electronvolts and the terahertz pump pulse with frequency 1 terahertz and field strength around $10$--$300$ kV/cm.
The typical size of QD is around $2$--$50$ nm, and this gives a rough estimation of the $h_0\!\sim\!0.02$--$3$ eV.

\begin{acknowledgement}
  We thank Han Htoon and Stuart Trugman for fruitful discussion.
  This work was carried out under the auspices of the U.S. Department of Energy (DOE) National Nuclear Security Administration under Contract No. 89233218CNA000001.
  It was supported by the Center for Integrated Nanotechnologies, a DOE Office of Science User Facility, and in part by the LANL LDRD Program.
  The computational resource was provided by the LANL Institutional Computing Program.

  The data that support the findings of this study are available from the corresponding author upon reasonable request.
\end{acknowledgement}

\bibliography{Zotero}

\end{document}




To calculate non-equilibrium tr-PL, we assume the system starts out in equilibrium at $t\!\rightarrow\!-\infty$ with Hamiltonian $H_\text{system}$.
In experiments, the spectrum may be measured by
\begin{equation}
  I_{{\bf e}_f}(t)=\sum_n\rho_n\langle\Psi^F_n\vert a^\dagger_{{\bf e}_f}a_{{\bf e}_f}\vert\Psi^F_n\rangle
\end{equation}
where $a^\dagger_{{\bf e}_f}$ is the photon creation operator with polarization ${\bf e}_f$
Typically, the pump pulse $H_\text{pump}$ is stronger than the probe pulse ${H}_\text{probe}$, and one can treat the probe pulse as a perturbation.
The full evolution operator can be approximated as

\begin{eqnarray}
 \hat{U}(t,t_0) &\simeq& U(t,t_0) -i\int^t_{t_0}dt_1 U(t,t_1)H_\text{probe}(t_1)U(t_1,t_0)\nonumber\\ &&+\int^t_{t_0}dt_2\int^{t_2}_{t_0}dt_1U(t,t_2)H_\text{probe}(t_2)U(t_2,t_1)H_\text{probe}(t_1)U(t_1,t_0)\;.
\end{eqnarray}
Here, the evolution operator is $U(t,t_0)\!=\!\mathcal{T}\exp[-i\int^t_{t_0}(H_\text{system}(t^\prime)+H_\text{pump}(t^\prime))dt^\prime)]$.
The leading contribution to the measured spectrum is of second order in $H_\text{probe}$ and we obtain
\begin{eqnarray}
  I_{{\bf e}_f}(t)&\simeq&\int^t_{t_0}dt_2\int^{t_2}_{t_0}dt_1\langle U(t,t_2){H}_\text{probe}(t_2)  U(t_2,\infty)[a^\dagger_{{\bf e}_f}a_{{\bf e}_f}-({\bf e}_f\cdot{\bf e}_i)^2] \nonumber\\ && U(\infty,t_1) {H}_\text{probe}(t_1)U(t_1,t_0)\rangle\;.
\end{eqnarray}
The probe pulse function is given at Eq.~(3) in the main text.
Putting all things together, we arrive
\begin{eqnarray}
  I_{{\bf e}_f}(t)&\simeq&\int^t_{t_0}dt_2\int^{t_2}_{t_0}dt_1\langle U(t,t_2){H}_\text{probe}(t_2)
   U(t_2,\infty)[a^\dagger_{{\bf e}_f}a_{{\bf e}_f}-({\bf e}_f\cdot{\bf e}_i)^2]U(\infty,t_1)
  \nonumber\\ && {H}_\text{probe}(t_1)U(t_1,t_0)\rangle\\
  &=&\sum_{{\bf e}_p,{\bf e}_p^\prime}\int^t_{t_0}dt_2\int^{t_2}_{t_0}dt_1s(t_1)s(t_2)\langle U(t,t_2)
   (A^\dagger_\text{probe}(t_2)a_{{\bf e}_p}+a^\dagger_{{\bf e}_p}A_\text{probe}(t_2))U(t_2,\infty)
  \nonumber\\ && [a^\dagger_{{\bf e}_f}a_{{\bf e}_f}-({\bf e}_f\cdot{\bf e}_i)^2]U(\infty,t_1)
  (A^\dagger_{m{\bf e}_p}(t_1)a_{{\bf e}^\prime_p}+a^\dagger_{{\bf e}_p}A_{m{\bf e}^\prime_p}(t_1))U(t_1,t_0)\rangle\;.\nonumber
\end{eqnarray}
Here, we further assume the tr-PL measurement takes an off resonance pump.
As mentioned before, we considered the system is in equilibrium in the past so the $t_0\!\rightarrow\!-\infty$.
Also, the time $t\rightarrow\infty$ is set so all the scattering events will be included.
Thus, the formula for tr-PL is
\begin{eqnarray}\label{eq:troas-final}
  I^\text{PL}(\omega_\text{probe})\simeq-\int^\infty_{-\infty}dt_2\int^\infty_{-\infty}dt_1e^{i\omega_\text{probe}(t_2-t_1)}s(t_1)s(t_2)C(t_1,t_2)\;.\nonumber
\end{eqnarray}
In equilibrium, the spectrum can be described as a continuous probe beam as $s(t)\!=\!1$.
In addition, the Hamiltonian is translational invarient in time and the ground state of the system is a vacuum state which allows us to take rotating wave approximation.
Therefore, the nonlocal time correlation function in Eq.~(5) of the maintext can be reduced to a function of $t\!=\!t_2-t_1$ as
\begin{eqnarray}
  &&I^\text{PL}(\omega_\text{probe})\propto
  -\sum_f\int^\infty_{-\infty}dte^{i(\omega_\text{probe}+E_i-E_f)t}\langle \psi_i \vert A_\text{probe} \vert \psi_f \rangle \nonumber \langle \psi_f \vert A^\dagger_\text{probe} \vert \psi_i \rangle\\
  &=&\sum_f \vert \langle \psi_f \vert A^\dagger_\text{probe} \vert \psi_i \rangle \vert^2 \delta(\omega_\text{probe}-E_i-E_f+i\Gamma)\;.\nonumber
\end{eqnarray}
Here, we usually consider the lifetime of the excitation is much shorter than the probe filed, thus an additional decay parameter ($\Gamma$) is introduced.

\begin{figure}[t]
  \begin{center}
    \includegraphics[width=0.6\columnwidth]{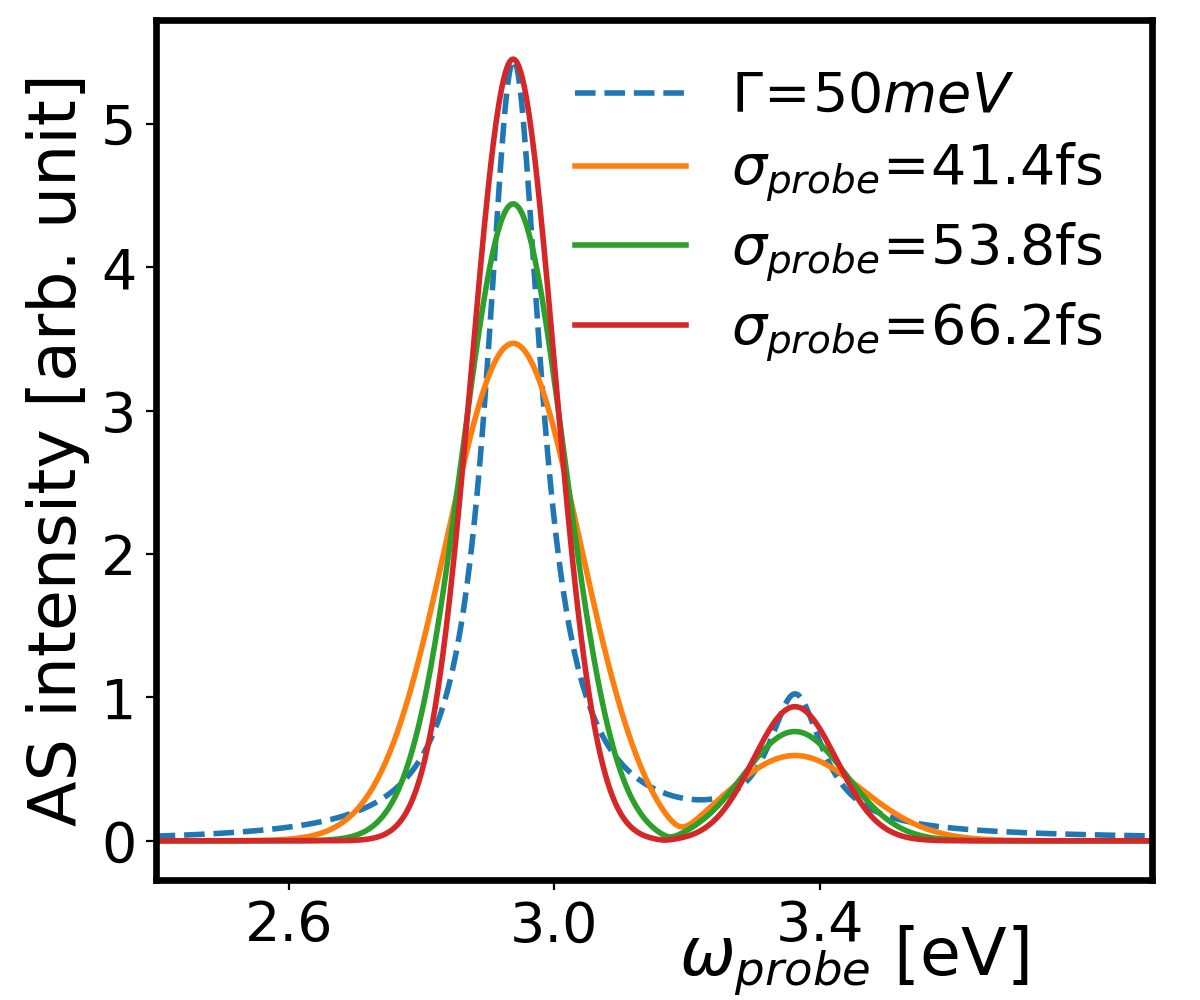}
    \caption{
      Absorption spectrum calculated from Fermi golden rule (dashed line) and nonlocal time correlation function (solid lines) with $E_d\!=\!3eV$, $E_c\!=\!3.3eV$, and $\Delta\!=\!150meV$.
    }
    \label{fig:Eqm}
  \end{center}
\end{figure}

Figure~\ref{fig:Eqm} show the comparison of PL calculated from the Fermi golden rule and the formula above in the absence of pump pulse.
The results clearly show that our formula recovers the equilibrium spectrum as increasing the width of the probe pulse and the spectrum becomes more broaden as the prob pulse becomes narrow.
